\documentclass[12pt,fullpage]{article}
\usepackage{amssymb,latexsym,amsmath,amscd,epsfig,amsthm,float}
\usepackage{color}
\input epsf
\newdimen\epsfxsize

\newtheorem{theorem}{Theorem}

\usepackage{setspace}
\begin{document}

\title{\large Adjusting for Misclassification: A Three-Phase Sampling Approach}
\author{Hailin Sang\footnote{Department of Mathematics, University of Mississippi,
University, MS 38677, USA. E-mail: sang@olemiss.edu},  Kenneth K. Lopiano \footnote{Roundtable Analytics, Research Triangle Park, NC 27709, USA. E-mail: klopiano@roundtableanalytics.com}, Denise A. Abreu,\\ Andrea C. Lamas \footnote{National Agricultural Statistics Service Research and Development, USDA, 1400 Independence Ave. SW, Room 6035,
USA. E-mail: Denise.Abreu@nass.usda.gov, Andrea.Lamas@nass.usda.gov}, Pam Arroway \footnote{EDUCAUSE, 282 Century Place, Ste 5000
Louisville, CO 80027, USA. E-mail: pam.arroway@gmail.com} and Linda J. Young \footnote{National Agricultural Statistics Service Research and Development, USDA, 1400 Independence Ave. SW, Room 6035,
USA. E-mail:  Linda.Young@nass.usda.gov} }
\date{}
\maketitle

\begin{abstract}
The United States Department of Agriculture's National Agricultural Statistics Service (NASS) conducts the June Agricultural Survey (JAS) annually. Substantial misclassification occurs during the pre-screening process and from field-estimating farm status for non-response and inaccessible records, resulting in a biased estimate of the number of US farms from the JAS. Here the Annual Land Utilization Survey (ALUS) is proposed as a follow-on survey to the JAS to adjust the estimates of the number of US farms and other important variables. A three-phase survey design-based estimator is developed for the JAS-ALUS with non-response adjustment for the second phase (ALUS). A design-unbiased estimator of the variance is provided in explicit form.

\textit{Keywords}: estimation under the three-phase sampling design,  non-response,  unbiased estimator, variance estimation
\end{abstract}

\section{Introduction}

\indent  The United States Department of Agriculture's National Agricultural Statistics Service (NASS) conducts numerous statistical surveys to provide information about current and future supplies of agricultural commodities.  See Fecso et al. (1986), Vogel (1995) and Nusser and House (2009) for the evolution and development of agricultural statistics and the surveys conducted at the United States Department of Agriculture. The June Agricultural Survey (JAS) is conducted annually. For the JAS, a stratified random sample is drawn using an area frame, which ensures complete coverage. Information about US crops, livestock, grain storage capacity, type and size of farms are collected from agricultural operations in the sample.  NASS uses the JAS to estimate numerous items relating to US agriculture, including the number of farms. 

Every five years, the annual number of farms estimate is compared to the one obtained from the quinquennial Census of Agriculture, which is a dual-frame survey conducted during years ending in 2 and 7.  See Kott and Vogel (1995) for details on the dual-frame survey. In 2007, the difference between the estimated number of farms from the JAS and the 2007 Census of Agriculture could not be attributed to sampling error alone. A preliminary study showed that the JAS estimate was biased because some farms were incorrectly classified as non-farms. In addition, some non-farms were misclassified as farms, but at a lower rate. Prior to this study, NASS had assumed that no misclassification was present in the JAS or any other survey that it conducted. 

Bross (1954) first showed that, when misclassification is present, conventional methods can be seriously biased. Tenenbein (1970, 1972) proposed a double-sampling scheme for inference from categorical data subject to misclassification. The double-sampling schemes utilize a sample of $n_1$ units classified by both a fallible and true device and another sample of $n_2$ units classified only by a fallible device. The double-sampling scheme and its variants are popular approaches to estimation when misclassification is present (see Thall et al., 1996, Stewart et al., 1998, and the references therein). Bayesian methods are also popular for inference from categorical data subject to misclassification (see Swartz et al., 2004,  the book by Gustafson, 2003, and the references therein). 

In this paper, a design-based approach that addresses misclassification and leads to improved estimates of the number of farms is suggested. First, the JAS sampling design is discussed, with an emphasis on the factors leading to the misclassification of farms and non-farms. Then, a proposed revision to the JAS sampling design is presented, and the properties of the resulting farm number estimates from this revised design explored. Finally, the implications of the work on the JAS are considered.

\section{The June Agricultural Survey (JAS)}
The JAS is conducted annually utilizing an area frame, ensuring complete coverage of the population. 
Land within the JAS area frame is divided into homogeneous land-use strata. Although minor definitional adjustments may be made depending on the specific needs of the state, land-use strata with more than 50\% cultivated land are generally labeled with a value in the 10s, agri-urban and commercial land-use strata are typically given a label in the 30s, etc. (see Table 1). The general land-use strata definitions are similar from state to state; however, minor definitional adjustments may be made depending on the specific needs of a state.
Each land-use stratum is further divided into substrata (called  \textquotedblleft design strata\textquotedblright) by grouping areas that are agriculturally similar, providing greater precision for state-level estimates of individual commodities. Within each design stratum, the land is divided into primary sampling units (PSUs). A sample of PSUs is selected and smaller, similar-sized segments (each of about a square mile (640 acres)) of land are delineated within these selected PSUs. Finally, one segment is randomly selected from each selected PSU to be fully enumerated. 

Once selected for inclusion in the JAS, a segment stays in the sample for five years. Thus, each year the sample has about 20\% new segments, and the 20\% of the segments that have been in the sample for five years rotate out. Segments rotating in during the same year are called a replicate; thus, each JAS sample consists of five replicates (see Cotter et al., 2010, Benedetti et al., 2015,  for further details on JAS). 

\begin{table}[H]
\center
\caption{ Land-use Strata} 
\label{tab1}
\bigskip
\begin{tabular}{| l | l |}
\hline                                                     
\;\;\;Land-use\;Strata  &  \\
\hline

\;\;\;$\ge$50\% \;cultivated  land&  \;\;\;\;10s\;\;\;\;\;\;\;\;\; \\
\hline

\;\;\;15-49\%\; cultivated land & \;\;\;\;20s \\
\hline

\;\;\;$<$15\% \; cultivated  land & \;\;\;\;40s \\
\hline

\;\;\;Agri-urban/Commercial  areas & \;\;\;\;30s \\
\hline

\;\;\;Non-agricultural land & \;\;\;\;50s \\
\hline
\end{tabular}  
\end{table}

Through 2010, the JAS pre-screening was conducted in the two weeks prior to data collection. During pre-screening, field enumerators (data collectors) divide each segment into tracts of land. Each tract represents a unique land operating arrangement. Field enumerators do not interview tract operators during pre-screening. Instead they complete an area screening form which provides an inventory of all tracts within a sampled segment and contains screening questions that determine whether or not each tract has agricultural activity. Using this form, each tract within the segment is screened for agricultural activity, and the screening applies to all land in the identified operating arrangement. Each screened tract is classified as agricultural or non-agricultural. Non-agricultural tracts are assigned to one of three categories: (1) non-agricultural with potential, (2) non-agricultural with unknown potential, or (3) non-agricultural with no potential. 

The JAS is conducted during the first two weeks of June. During the sampling period, field enumerators return to {\it only} those tracts classified as agricultural during the earlier screening period.  Data collection continues until some type of response is obtained for every sampled tract. If a respondent cannot be reached, the information may be obtained from administrative data, data collected for other surveys, or estimates made by field enumerators. Regardless of the information source, these tracts are identified as being field estimated.  Based on the JAS, an agricultural tract is classified as a farm if its entire operation, which could include land outside the sampled tract, qualifies with at least \$1,000 in agricultural sales or potential sales. All non-agricultural tracts and agricultural tracts with less than \$1,000 in sales are classified as non-farms. 

In 2009, NASS conducted a one-time follow-on survey to the JAS segments, the Farm Numbers Research Project (FNRP) (Abreu et al.  2010). The sampling design of the FNRP targeted the 20\% of JAS segments that were newly rotated in for 2009 (2009 segments). All tracts in the 2009 segments that were non-agricultural or field estimated in JAS were selected for FNRP. 
During the FNRP,
all places of interest within a selected tract were considered subtracts. 

A shortened form based on
the JAS questionnaire was used to classify each subtract as a farm or a non-farm.

A major finding in FNRP was that, assuming misclassification rates are the same for all rotations (did not differ from that observed for the 2009 segments), the JAS estimate of number of farms would increase by approximately 580,000 farms (see Table 2). The bulk of these farms were ÒfoundÓ in tracts that had been identified as non-agricultural with no potential in the JAS.

\begin{table}[H]
\center
\caption{FNRP Results by Type of Tract } 
\label{tab2}
\bigskip
\begin{tabular}{| l | l | l | l |}
\hline                                                    
Type of Tract   &  FNRP Sample & Number of  & Net Expanded \\
                        &Size (subtracts) & FNRP Farms                & Number of Farms \\
\hline

Field estimated    &  1,591  & 1,466  & (7,822) \\
as farm  &   &    &\\
\hline

Field estimated     &  121  & 37  & 13,032 \\
as non-farm  &   &    &\\
\hline

Non-agricultural    &  487  & 95  & 38,346\\
with potential   &   &    &\\
\hline

Non-agricultural      &  364  & 56  & 37,479 \\
with unknown    &   &    &\\
potential            &   &     &\\
\hline

Non-agricultural      &  14,628   & 905  & 500,338  \\
with no potential   &   &    &\\
\hline
FNRP Total    &  17,191   & 2,559  & 581,373  \\
\hline
\end{tabular}  
\end{table}

Several factors could lead to the misclassification of farms as non-farms and non-farms as farms. During pre-screening, the agricultural activity may not have been evident when the field enumerator observed the tract from a distance (tract operators are not interviewed during this process), or the primary agricultural activity could have been outside of the sampled tract (the response for a tract includes agriculture associated with {\it all} of the operation, not just that within the tract). In FNRP, 86.1 percent (500,338) of the field estimated number of farms misclassified as non-farms were found in tracts pre-screened to be non-agricultural with no potential. Small farms are more likely to be misclassified. In FNRP, 58.3 percent (335,902) of the field estimated number of farms misclassified as non-farms had less than 25 acres. Operations that recently went out of business or small farms whose production fell below the \$1000 threshold in sales could be misclassified as farms when field estimated.  

To obtain a more accurate estimate of the number of US farms from the JAS, the current estimation approach must be revised to account for misclassification. The Annual Land Utilization Survey (ALUS), a follow-on survey to the JAS, has been proposed for this purpose. FNRP results are used as guidelines for the ALUS design, but ALUS will be able to detect different types of trends as well.

\section{The Annual Land Utilization Survey (ALUS): Design}
The ALUS focuses on those JAS tracts that were potentially misclassified as farm or non-farms either during the pre-screening process or during field estimation of farm status for non-responding or inaccessible operations. These tracts are treated as non-responders, and data collection is focused on obtaining accurate information on them. ALUS represents the second phase of a two-phase sample with the first phase being the traditional JAS. As in the JAS, the proposed ALUS is a stratified sample of segments, using JAS land-use strata and sampling across rotations. Segments that are eligible for inclusion in ALUS must have at least one tract that was pre-screened as non-agricultural (regardless of potential) or that was field estimated in JAS (as either a farm or non-farm); that is, only JAS segments that had completed interviews for all tracts are {\it not} eligible for possible inclusion in the ALUS sample. For a selected segment, all tracts are to be re-evaluated using a modified combined JAS-ALUS questionnaire. The collection of eligible segments in a particular year will be called the ALUS population.

For ALUS, the sample allocation of segments to each state-stratum combination considers two factors: the proportion of the ALUS population in the land-use stratum and the proportion of the FNRP adjustment from non-agricultural tracts in the land-use stratum (see Table 3). The latter simultaneously accounts for the number of converted non-agricultural tracts and the expansion factors associated with them, allowing states and land-use strata that contributed most to the FNRP adjustment to be targeted. In the JAS, the sampling scheme favors cultivated areas. For ALUS, the sampling will lean more heavily on moderately and less cultivated land-use strata where the largest portion of the FNRP adjustment originates. For example, although the exact land-use stratum definition varies from state to state, land-use strata 10s $(10, 11, \cdots)$ are highly cultivated areas, with generally at least 50\% cultivated land. In the JAS, over half of the selected segments are from these land-use strata. However, 10s made up only 16\% of the FNRP adjustment arising from non-agricultural tracts so only about 27\% of the ALUS sample will come from these strata. The sample will be evenly distributed over the five rotations, with approximately 20\% of the ALUS sample selected from each. 

\begin{table}[H]
\center
\caption{Guidelines for ALUS Allocation Scheme} 
\label{tab3}
\bigskip
\begin{tabular}{| l | l | l | l | l |}
\hline                                                      
Land-    & Proportion   & Proportion    & Proportion   &  Suggested \\
   use             &of FNRP  &  of ALUS-             & of ALUS-            & Proportion \\
   strata                     & adjustment  &  eligible              & eligible           & of ALUS\\
                        & from non-  &  segments in             & segments in           & sample\\
                        & agricultural  &  2009 JAS              & 2010 JAS               & \\
                         & tracts          &                                &                               &\\ 
\hline

10s   &  16\%  & 53\%  & 52\%  & 27\% \\
\hline

20s   &  34\%  & 26\%  & 27\%  & 30\% \\
\hline

30s   &  $<$1\%  & 3\%  & 3\%  & 3\% \\
\hline

40s   &  50\%  & 17\%  & 17\%  & 39\% \\
\hline

50s   &  $<$1\%  & $<$1\%  & $<$1\%  & 1\% \\
\hline
Total   &  576,000 farms   & 10,168 segments   & 10,121 segments   &  \\
\hline
\end{tabular}  
\end{table}

Within each land-use stratum of the ALUS population, segments will be selected with probability
proportional to size (pps) sampling where the size measure of a segment is defined as the sum of the number of tracts either pre-screened as non-agricultural or field estimated to be non-farms and one-tenth of the number of tracts  field estimated to be a farm. Because most tracts (92\%)  field estimated as farms in the JAS were confirmed as farms in FNRP, ALUS only takes a tenth of the number of these tracts within a segment when determining size. If a segment is selected, all ALUS-eligible tracts within that segment will be in the sample, including those  field estimated as farms.

Precise estimates of uncertainty can be obtained by viewing the combination of JAS and ALUS as a two-phase sample, with JAS being the first phase and ALUS being the second. Given that each phase makes use of a probability sampling design with known inclusion probabilities, standard results can be used to construct a design-based estimator (S\"{a}rndal and Swensson, 1987). However, non-response is also expected to occur in ALUS. Instead of using the estimated tract values to account for this non-response, the two-phase design estimator of S\"{a}rndal and Swensson (1987) has been extended to a third phase (see Section 4). The resulting estimator is used for the two-phase JAS-ALUS, with the self-selection of response treated as a third phase of random sampling.  This methodology can be applied not only to estimates of the number of farms but to all variables collected in the ALUS.

\section{Estimation} 
In this section we first extend the two-phase $\pi^*$  estimator (S\"{a}rndal and Swensson, 1987) 
to three-phase survey sampling estimator. Legg and Fuller (2009),  S\"{a}rndal et al. (1992) and Singh (2003) provide a review of the two-phase sampling estimator. Jeyaratnam et al. (1984) studied a multiphase design in a forest study. Fuller (2003) studied a three-phase regression estimator for the mean of a vector population. Magnussen (2003) studied estimators for three-phase sampling of categorical variables.  Then in the second subsection we study the application to the ALUS estimator with non-response adjustment. 

\subsection{Estimation under a three-phase sampling design}
\noindent 
To be consistent and complete, the notation used by S\"{a}rndal and Swensson (1987) for the two-phase design is extended for the third phase.  

Let $y_k$  be the response of interest for the $k^{th}$ unit in a finite population $U$. The population total is $T=\sum_U y_k$. A general sampling design is allowed in each phase.

(a) The first-phase sample $S (S\subset U)$  is drawn according to a sampling design $P_a(\cdot)$, such that  $P_a(S)$  is the probability of choosing S. The inclusion probabilities are defined by 
$$\pi_{ak}=\sum_{k\in S}P_a(S),  \pi_{akp}=\sum_{k,p\in S}P_a (S)$$
with $\pi_{akk}=\pi_{ak}$.  Set $\Delta_{akp}=\pi_{akp}-\pi_{ak}\pi_{ap}$.   It is assumed that $\pi_{ak}>0$ for all $k$, $\pi_{akp}>0$ for all $k\ne p$ in variance estimation. $\pi_{ak}$ is the probability of selection of the $k^{th}$ unit in the first phase sampling. $\pi_{akp}$   is the probability of selection both the $k^{th}$ unit and the $p^{th}$  unit in the first phase sampling.

(b) Given $S$,  the second-phase sample $R (R\subset S)$ is drawn according to a sampling design  $P(\cdot |S)$,  such that  $P(R|S)$, is the conditional probability of choosing $R$.  The inclusion probabilities given $S$ are defined by 
$$\pi_{k|S}=\sum_{k\in R}P(R|S),  \pi_{kp|S}=\sum_{k,p \in R}P(R|S).$$ 
$\pi_{kk|S}=\pi_{k|S}$. Set $\Delta_{kp|S}=\pi_{kp|S}-\pi_{k|S} \pi_{p|S}$.  It is assumed that for any $S$, $\pi_{k|S}>0$ for all $k\in S$, $\pi_{kp|S}>0$ for all $k\ne p\in S$ in variance estimation.  $\pi_{k|S}$ is the probability of selection of the $k^{th}$ unit in the second phase sampling given the result of the first phase sampling. $\pi_{kp|S}$ is the probability of selecting both the $k^{th}$  unit and the $p^{th}$ unit in the second phase sampling given the result of the first phase sampling.

(c) Given $R$,  the third-phase sample $F (F\subset R)$ is drawn according to a sampling design $P(\cdot |R)$,  such that  $P(F|R)$, is the conditional probability of choosing $F$.  $F$  is the set of selected units in a three-phase sampling design or the set of responses for the second phase in a two-phase sampling design. The inclusion probabilities given $R$ are defined by 
$$\pi_{k|R}=\sum_{k\in F} P(F|R), \pi_{kp|R}=\sum_{k,p\in F}P(F|R).$$  
$\pi_{kk|R}=\pi_{k|R}$. Set $\Delta_{kp|R}=\pi_{kp|R}-\pi_{k|R} \pi_{p|R}$. In a three-phase sampling design, $\pi_{k|R}$ is the probability of selection of the $k^{th}$ unit in the third phase of sampling given the result of the first two phases of sampling. $\pi_{kp|R}$ is the probability of selecting both the $k^{th}$  unit and the $p^{th}$ unit in the third phase of sampling given the result of the first two phases of sampling.  In a two-phase sampling design, $\pi_{k|R}$  is the probability when the $k^{th}$ unit has response for the second phase.  $\pi_{kp|R}$ is the probability that both the $k^{th}$ unit and the $p^{th}$  unit have a response for the second phase.

Now for any $S$ and for all $k,p\in S$, define $\pi_k^*=\pi_{ak} \pi_{k|S}$, $\pi_{kp}^*=\pi_{akp} \pi_{kp|S}$. $\pi_{kk}^*=\pi_k^*$.    
Next, define
$\pi_k^{\#}=\pi_k^* \pi_{k|R}=\pi_{ak}\pi_{k|S}\pi_{k|R}$
for all $k\in R$  and any $R$.
Then the first-phase expanded $y$-value is $\breve{y}_k=y_k/\pi_{ak}$.  The second-phase expanded $y$-value is $\breve{y}_k^*=\breve{y}_k/\pi_{k|S} =y_k/\pi_k^*$.  The third-phase expanded $y$-value  is $\breve{y}_k^{\#}=\breve{y}_k^*/\pi_{k|R} =\breve{y}_k/(\pi_{k|S}\pi_{k|R})=y_k/(\pi_{ak}\pi_{k|S} \pi_{k|R} )=y_k/\pi_k^{\#}$.  The expanded $\Delta$ values are $\breve{\Delta}_{akp}=\Delta_{akp}/\pi_{akp}$, $\breve{\Delta}_{kp|S}^*=\Delta_{akp}/(\pi_{kp}^*)=\Delta_{akp}/(\pi_{akp} \pi_{kp|S})$. $\breve{\Delta}_{kp|S}=\Delta_{kp|S}/\pi_{kp|S}$.  
Now, the expansion  estimator in three-phase sampling is defined as, 
\begin{equation}\label{est}
\hat{t}_{\#}=\sum_{k\in F}\breve{y}_k^{{\#}} =\sum_{k\in F}y_k/\pi_k^{\#}.
\end{equation}
The following theorem gives an unbiased estimator of the variance of the triple expansion estimator $\hat{t}_{\#}$. 
\begin{theorem}\label{maintheorem}
 The estimator in (\ref{est}) is design unbiased, and a design unbiased estimator of $Var(\hat{t}_{\#})$ is given by
\begin{eqnarray}\label{varest}
\widehat{Var}(\hat{t}_{\#})=\sum\sum_F\breve{\Delta}_{kp|S}^* \breve{y}_k \breve{y}_p/\pi_{kp|R}\notag\\
+\sum\sum_F\breve{\Delta}_{kp|S} \breve{y}_k^* \breve{y}_p^*/\pi_{kp|R}+\sum\sum_F\Delta_{kp|R} \breve{y}_k^{\#} \breve{y}_p^{\#}/\pi_{kp|R}.       \end{eqnarray}
\end{theorem}
The proof of Theorem \ref{maintheorem} is deferred to Appendix. 
\subsection{The ALUS estimator}
Let $T$ be the United States number of farms in a specific year. First, consider the JAS estimate of the number of farms. Then the estimator incorporating the information obtained during the ALUS (second-phase sample) and the non-response adjustment in ALUS will be developed.

Under stratified simple random sampling, the JAS estimator of $T$ is
	\begin{equation}\label{jas}
	\hat{T}=\sum_{i=1}^l\sum_{j=1}^{s_i}d_{ij}\sum_{k=1}^{n_{ij}}\sum_{m=1}^{x_{ijk}}t_{ijkm}
	\end{equation}
where 	
\begin{itemize}
	\item $i$ is the index of land-use stratum,  $l$ is the number of land-use strata;
	\item $j$ is the index of design stratum,  $s_i$ is the number of design strata in land-use stratum $i$;
	\item $k$ is the index of segment,  $n_{ij}$ is the number of segments in design stratum $j$ within land-use stratum $i$;
	\item $d_{ij}$ is the expansion factor or the inverse of the probability of selection for each segment in design stratum $j$ in land-use stratum $i$;
	\item $m$ is the index of tract,  $x_{ijk}$ is the number of {\it farm} tracts in the segment;
	\item $t_{ijkm}$ is the tract-to-farm ratio, which is $\frac{tract\;\; acres\;\; for\;\; the\;\; m^{th} \;\;tract}{farm\;\; acres\;\; for\;\; the\;\; m^{th}\;\; tract}$.
\end{itemize}

Under the assumption that the JAS provides accurate information for all tracts, $\hat{T}$ is unbiased.
The variance is
	\begin{equation}\label{JASt1}
	Var(\hat{T})=\sum_{i=1}^l\sum_{j=1}^{s_i}\frac{1-1/d_{ij}}{1-1/n_{ij}}\sum_{k=1}^{n_{ij}}(c_{ijk}-c_{ij.})^2
	\end{equation}
where $c_{ijk}=d_{ij}\sum_{m=1}^{x_{ijk}}t_{ijkm}$,  $c_{ij.}=\frac{1}{n_{ij}}\sum_{k=1}^{n_{ij}}c_{ijk}$. This formula is given by Kott (1990).

However, the JAS estimate is biased because some tracts are misclassified either during pre-screening when agricultural tracts may be identified as non-agricultural or during the JAS when tracts are incorrectly field estimated to be farms or non-farms. 

Now consider the JAS-ALUS two-phase estimator with nonresponse adjustment for the second phase. The estimator is:
\begin{equation}\label{alus2}
	\hat{\hat{T}}=\hat{T_1}+\sum_{i=1}^{l}\sum_{j=1}^{s_i}d_{ij}a_{ij}\sum_{k=1}^{n_{ij}'}r_{ijk}\sum_{m=1}^{z_{ijk}}t_{ijkm}:=\hat{T_1}+\hat{T_2}.
	\end{equation}
Here the first term $\hat{T_1}$ has the same form as $\hat{T}$ in (\ref{jas}). But,  it only includes the JAS segments comprised of all farm tracts confirmed through an interview of the operator (not estimated) in the first phase. In the second phase, the ALUS sample only includes the JAS tracts that were either pre-screened as non-agricultural  or field estimated as either a farm or a non-farm. Thus each tract in the ALUS sample has potentially been misclassified and is treated as a non-respondent from the first phase.  $n_{ij}'$ is the number of ALUS segments in design stratum $j$ within land-use stratum $i$. $a_{ij}$ is the expansion factor or the inverse of the probability of selection in the second phase for each segment in design stratum $j$ in land-use stratum $i$. $z_{ijk}$ is the number of farm tracts in the given ALUS selected segment.  $r_{ijk}$ is the expansion factor or the inverse of the response probability of each tract  in segment $k$, design stratum $j$, land-use stratum $i$. 

Here we assume that all tracts in the same segment have the same response probability and this probability $r_{ijk}$ is known. If $r_{ijk}$ is unknown, it can be estimated by modeling under the assumption of stratified Bernoulli subsampling for non-response, i.e., a response is assumed to have the Bernoulli distribution. In this case, we would have another variance component. This is a complex case and is not considered here.  A referee suggested that, instead of assuming $r_{ijk}$ known, the last phase could be treated conditionally (on the number of good responses) as a simple random sample within each segment. The assumption needed for this approach is at least two responses are obtained within each segment. Readers are referred to S\"{a}rndal et al.  (1992) for the modeling on non-response in a quasi-design-based framework (\textquotedblleft quasi\textquotedblright  because response if modeled). Hidiroglou and Estevao (2013) used a follow-up sample of the non-respondents to deal with nonresponse.

Now we apply (\ref{varest}) in Theorem \ref{maintheorem} to obtain a design-unbiased estimator of $Var(\hat{T_2})$. For convenience, we use $(i, j)$ to denote design stratum $j$ within land-use stratum $i$. We also use $k$ or $p$ to be the index of segment. 
In the JAS-ALUS sampling design, the unit is a segment. One unit is one segment in $(i, j)$. It includes all tracts in that segment. Recall that all segments within the same design stratum have the same expansion factor. The first phase expansion factor is $d_{ijk}=d_{ij}$ and the second phase expansion factor is $a_{ijk}=a_{ij}$ for all segments $k$ in $(i, j)$. Therefore, $\pi_{ak}=d_{ijk}^{-1}=d_{ij}^{-1}$,  and 
\begin{equation}\label{1sty}
\breve{y}_k=y_k/\pi_{ak} =d_{ij}\sum_{m=1}^{z_{ijk}}t_{ijkm}.
\end{equation} 
$\pi_{k|S}=a_{ij}^{-1}$. There are $n_{ij}'a_{ij}d_{ij}$ segments in $(i, j)$.  If $k\ne p$ and these segments are in a same design stratum $(i, j)$, 
$$\pi_{akp}=(n_{ij}'a_{ij}-1)/[d_{ij}(n_{ij}'a_{ij}d_{ij}-1)],$$ 
$$\Delta_{akp}=\pi_{akp}-\pi_{ak} \pi_{ap}=(1-d_{ij})/[d_{ij}^2(n_{ij}'a_{ij} d_{ij}-1)],$$  
$$\breve{\Delta}_{akp}=\Delta_{akp}/\pi_{akp} =(1-d_{ij})/[d_{ij}(n_{ij}'a_{ij}-1)],$$  
$$\pi_{kp|S}=(n_{ij}'-1)/[a_{ij} (n_{ij}'a_{ij}-1)].$$
If $k,p$ are from different design stratum $(i,j),(i',j')$,  $\Delta_{akp}=0.$ $\breve{\Delta}_{akp}=0$.  $\pi_{kp|S}=1/(a_{ij} a_{i' j'})$. 
If $k=p$,  
$$\pi_{akk}=\pi_{ak}=d_{ij}^{-1}, \Delta_{akk}=d_{ij}^{-1}-d_{ij}^{-2},$$ 
$$\breve{\Delta}_{akk}=\Delta_{akk}/\pi_{akk}=1-d_{ij}^{-1}, \pi_{kk|S}=\pi_{k|S}=a_{ij}^{-1}.$$
Therefore, 
$$\breve{\Delta}_{kp|S}^*=\breve{\Delta}_{akp}/\pi_{kp|S} =a_{ij} (1-d_{ij})/[d_{ij}(n_{ij}'-1)]$$ 
if $k\ne p$ are in the same design stratum. 
$\breve{\Delta}_{kp|S}^*=0$ 
if $k ,p$ are from different design stratum. 
\begin{equation}					
\breve{\Delta}_{kp|S}^*=\breve{\Delta}_{akp}/\pi_{kp|S} =[a_{ij} (d_{ij}-1)]/d_{ij}                                                                  
\end{equation}
if $k=p$. 
In the second phase of ALUS, recall that $\pi_{k|S}=\pi_{kk|S}=1/a_{ij}$,  $\pi_{kp|S}=(n_{ij}'-1)/[a_{ij} (n_{ij}'a_{ij}-1)]$   if the two different segments are in the same design stratum. Otherwise, $\pi_{kp|S}=1/(a_{ij} a_{i'j'})$.  Therefore, 
\begin{equation}\label{breveY}	
\breve{y}_k^*=\breve{y}_k/\pi_{k|S} =d_{ij} a_{ij}\sum_{m=1}^{z_{ijk}}t_{ijkm}.
\end{equation}
$$\Delta_{kp|S}=\pi_{kp|S}-\pi_{k|S}\pi_{p|S}=(1-a_{ij})/[a_{ij}^2 (n_{ij}'a_{ij}-1)]$$
 and $$\breve{\Delta}_{kp|S}=\Delta_{kp|S}/\pi_{kp|S}=(1-a_{ij})/[a_{ij}(n_{ij}'-1)]$$
  if the two different segments are in the same design stratum. $\Delta_{kp|S}=0=\breve{\Delta}_{kp|S}$ if the two segments are in different design stratum. $\Delta_{kp|S}=\pi_{kp|S}-\pi_{k|S} \pi_{p|S}=(a_{ij}-1)/(a_{ij}^2)$  and $\breve{\Delta}_{kp|S}=\Delta_{kp|S}/\pi_{kp|S} =(a_{ij}-1)/a_{ij}$    if $k=p$.  $\pi_{k|R}$ is the probability of response of the tracts in segment $k$.  $\pi_{kp|R}$ is the probability that two tracts have response in segments $k, p$. $\pi_{k|R}=\pi_{kk|R}=1/r_{ijk}$ and $\pi_{kp|R}=1/(r_{ijk}r_{ijp})$ if $k\ne p$. Then $\Delta_{kp|R}=\pi_{kp|R}-\pi_{k|R} \pi_{p|R}=0$ if $k\ne p$ and $\Delta_{kk|R}=\pi_{k|R}-\pi_{k|R}^2=(r_{ijk}-1)/r_{ijk}^2$.  By (\ref{breveY}), the third-phase expanded $y$-value 
  $$\breve{y}_k^{\#}=\breve{y}_k^*/\pi_{k|R}=d_{ij} a_{ij}r_{ijk}\sum_{m=1}^{z_{ijk}}t_{ijkm}.$$  
Together with all the analysis, the design-unbiased estimator (\ref{varest}) of $Var(\hat{T_2})$ is  
\begin{align}
&\widehat{Var}(\hat{T_2})=\sum_{i=1}^l\sum_{j=1}^{s_i}a_{ij}d_{ij}(d_{ij}-1)\sum_{k=1}^{n_{ij}'}r_{ijk}(\sum_{m=1}^{z_{ijk}}t_{ijkm})^2\notag\\
&+\sum_{i=1}^l\sum_{j=1}^{s_i}d_{ij} a_{ij}(1-d_{ij})(n_{ij}'-1)^{-1}\sum_{1\le k<p\le n_{ij}'}(\sum_{m=1}^{z_{ijk}}r_{ijk}t_{ijkm}\sum_{m=1}^{z_{ijp}}r_{ijp}t_{ijpm})\notag\\ 
&+\sum_{i=1}^l\sum_{j=1}^{s_i}d_{ij}^2 a_{ij}(a_{ij}-1)\sum_{k=1}^{n_{ij}'}r_{ijk}(\sum_{m=1}^{z_{ijk}}t_{ijkm})^2\notag\\                 
&+\sum_{i=1}^l\sum_{j=1}^{s_i}d_{ij}^2 a_{ij}(1-a_{ij})(n_{ij}'-1)^{-1}\sum_{1\le k<p\le n_{ij}'}(\sum_{m=1}^{z_{ijk}}r_{ijk}t_{ijkm}\sum_{m=1}^{z_{ijp}}r_{ijp}t_{ijpm})\notag\\    
&+\sum_{i=1}^l\sum_{j=1}^{s_i}d_{ij}^2 a_{ij}^2 \sum_{k=1}^{n_{ij}'}r_{ijk}(r_{ijk}-1)(\sum_{m=1}^{z_{ijk}}t_{ijkm})^2 .\label{t2var}
\end{align}     
In (\ref{t2var}), the first two summands give the first quantity in (\ref{varest}); summand $3$ and $4$ give the second quantity in (\ref{varest}); and the last summand gives the third quantity in (\ref{varest}). $\widehat{Var}(\hat{T_2})$ can be further simplified to 
\begin{eqnarray}\label{variance2}
\widehat{Var}(\hat{T_2})=\sum_{i=1}^l\sum_{j=1}^{s_i}\sum_{k=1}^{n_{ij}'}a_{ij}d_{ij}r_{ijk}(a_{ij}d_{ij}r_{ijk}-1)(\sum_{m=1}^{z_{ijk}}t_{ijkm})^2\notag\\
+\sum_{i=1}^l\sum_{j=1}^{s_i}d_{ij} a_{ij}(1-d_{ij}a_{ij})(n_{ij}'-1)^{-1}\sum_{1\le k<p\le n_{ij}'}(\sum_{m=1}^{z_{ijk}}r_{ijk}t_{ijkm}\sum_{m=1}^{z_{ijp}}r_{ijp}t_{ijpm}).
\end{eqnarray}    
We denote $\widehat{Var}(\hat{T_2})=\sum_{i=1}^l\sum_{j=1}^{s_i}V_{ij}$ where $V_{ij}$ is the contribution to the variance from the segments in design stratum $j$ in land-use stratum $i$. In the special case that $r_{ijk}=r_{ijp}=r_{ij}$ and $\sum_{m=1}^{z_{ijk}}t_{ijkm}=\sum_{m=1}^{z_{ijp}}t_{ijpm}=c_{ij}$, $1\le k<p\le n_{ij}'$, for some $i,j$, the $V_{ij}$ is 
\begin{eqnarray}
V_{ij}=n_{ij}'a_{ij}d_{ij}r_{ij}(a_{ij}d_{ij}r_{ij}-1)c_{ij}^2\notag\\
+d_{ij} a_{ij}(1-d_{ij}a_{ij})(n_{ij}'-1)^{-1}r_{ij}^2\frac{n_{ij}'(n_{ij}'-1)}{2}c_{ij}^2\notag\\
=\frac{1}{2}d_{ij}a_{ij}r_{ij}n_{ij}'[r_{ij}(d_{ij}a_{ij}+1)-2]c_{ij}^2.
\end{eqnarray}    
$V_{ij}\ge 0$ as expected since the expansion factors $d_{ij}, a_{ij}, r_{ij}\ge 1$. The contribution $V_{ij}=0$ if $d_{ij}=a_{ij}=r_{ij}=1$. $\widehat{Var}(\hat{T_2})=0$ if $d_{ij}=a_{ij}=r_{ij}=1$ for all $i, j$. This is the case of complete census without non-response. 

To derive the variance of $\hat{\hat{T}}$, let $E(\cdot |JAS)$ and $Var(\cdot |JAS)$ refer, respectively, to the conditional expectation and conditional variance given the outcome of the JAS. We use the formula 
\begin{align}\label{v}
Var(\hat{\hat{T}})&=Var(\hat{T_1}+\hat{T_2})\nonumber\\
&=E[Var(\hat{T_1}+\hat{T_2}|JAS)]+Var[E(\hat{T_1}+\hat{T_2}|JAS)]\nonumber\\
&=E[Var(\hat{T_2}|JAS)]+Var[\hat{T_1}+E(\hat{T_2}|JAS)].
\end{align}
By the proof of Theorem \ref{maintheorem}, the first term of (\ref{v}) is estimated by the second and third quantities in Theorem \ref{maintheorem}, which are the summands 3, 4 and 5 in (\ref{t2var}).  
By (\ref{conExp1}) in the Appendix and (\ref{1sty}), 
\begin{align*}
E(\hat{T_2}|JAS)=\sum_{i=1}^{l}\sum_{j=1}^{s_i}d_{ij}\sum_{k=1}^{a_{ij}n_{ij}'}\sum_{m=1}^{z_{ijk}}t_{ijkm}.
\end{align*}
Here  $a_{ij}n_{ij}'$ is the number of segments in the ALUS population in design stratum $j$ within land-use stratum $i$ since $a_{ij}$ is the expansion factor and $n_{ij}'$ is the number of ALUS segments in $(i, j)$. Together with (\ref{jas}), we have 
	\begin{equation*}
	\hat{T_1}+E(\hat{T_2}|JAS)=\sum_{i=1}^l\sum_{j=1}^{s_i}d_{ij}\left(\sum_{k=1}^{n_{ij}}\sum_{m=1}^{x_{ijk}}t_{ijkm}+\sum_{k=1}^{a_{ij}n_{ij}'}\sum_{m=1}^{z_{ijk}}t_{ijkm}\right).
	\end{equation*}
By	(\ref{JASt1}), 
\begin{align}\label{vart1et2}
&Var[\hat{T_1}+E(\hat{T_2}|JAS)]\nonumber\\
&=	\sum_{i=1}^l\sum_{j=1}^{s_i}\frac{1-1/d_{ij}}{1-1/(n_{ij}+a_{ij}n_{ij}')}\sum_{k=1}^{n_{ij}+a_{ij}n_{ij}'}(c_{ijk}-c_{ij.})^2
\end{align}
where $c_{ijk}=d_{ij}\sum_{m=1}^{x_{ijk}}t_{ijkm}$, if $1\le k\le n_{ij}$, $c_{ijk}=d_{ij}\sum_{m=1}^{z_{ijk}}t_{ijkm}$, if $n_{ij}+1\le k\le n_{ij}+a_{ij}n_{ij}'$,  $c_{ij.}=\frac{1}{n_{ij}+a_{ij}n_{ij}'}\sum_{k=1}^{n_{ij}+a_{ij}n_{ij}'}c_{ijk}$. 

Nevertheless, we cannot calculate (\ref{vart1et2}) since only the ALUS sample information, which includes $n_{ij}'$ segments in $(i, j)$, is known. A design unbiased estimator of  (\ref{vart1et2}) is given by
\begin{align}
&\widehat{Var}[\hat{T_1}+E(\hat{T_2}|JAS)]\\
&=	\sum_{i=1}^l\sum_{j=1}^{s_i}\frac{1-1/d_{ij}}{1-1/(n_{ij}+a_{ij}n_{ij}')}\left(\sum_{k=1}^{n_{ij}}(c_{ijk}-\widehat{c_{ij.}})^2+a_{ij}\sum_{p=1}^{n_{ij}'}(\widehat{c_{ijp}}-\widehat{c_{ij.}})^2\right),\nonumber
\end{align}
where $\widehat{c_{ijp}}=d_{ij}r_{ijp}\sum_{m=1}^{z_{ijp}}t_{ijpm}$, $1\le p\le n_{ij}'$,   and $$\widehat{c_{ij.}}=\frac{1}{n_{ij}+a_{ij}n_{ij}'}\left(\sum_{k=1}^{n_{ij}}c_{ijk}+a_{ij}\sum_{p=1}^{n_{ij}'}\widehat{c_{ijp}}\right).$$ 
Hence we have the design unbiased estimator of  $Var(\hat{\hat{T}})$, 
\begin{align}
&\widehat{Var}(\hat{\hat{T}})=\sum_{i=1}^l\sum_{j=1}^{s_i}d_{ij}^2 a_{ij}(a_{ij}-1)\sum_{k=1}^{n_{ij}'}r_{ijk}(\sum_{m=1}^{z_{ijk}}t_{ijkm})^2\notag\\                 
&+\sum_{i=1}^l\sum_{j=1}^{s_i}d_{ij}^2 a_{ij}(1-a_{ij})(n_{ij}'-1)^{-1}\sum_{1\le k<p\le n_{ij}'}(\sum_{m=1}^{z_{ijk}}r_{ijk}t_{ijkm}\sum_{m=1}^{z_{ijp}}r_{ijp}t_{ijpm})\notag\\    
&+\sum_{i=1}^l\sum_{j=1}^{s_i}d_{ij}^2 a_{ij}^2 \sum_{k=1}^{n_{ij}'}r_{ijk}(r_{ijk}-1)(\sum_{m=1}^{z_{ijk}}t_{ijkm})^2 \notag\\
&+\sum_{i=1}^l\sum_{j=1}^{s_i}\frac{1-1/d_{ij}}{1-1/(n_{ij}+a_{ij}n_{ij}')}\left(\sum_{k=1}^{n_{ij}}(c_{ijk}-\widehat{c_{ij.}})^2+a_{ij}\sum_{p=1}^{n_{ij}'}(\widehat{c_{ijp}}-\widehat{c_{ij.}})^2\right).\nonumber
\end{align}
\section{Conclusions}
The JAS is the largest annual survey conducted by NASS. Its results are
used to develop a number of official estimates. Here the focus has been on
estimating the number of US farms. The substantial misclassification of
farms and non-farms has led
to a biased estimate of the number of farms. The two-phase JAS-ALUS has been suggested as an improvement that would produce a (quasi-)unbiased estimation
of farm numbers. The proposed three-phase survey design-based estimator
 (\ref{est}) is an extension of the two-phase sampling estimator in  S\"{a}rndal and Swensson (1987), which allows for a general sampling design in each phase.
For the JAS-ALUS application considered here, the JAS is the first phase;
ALUS is the second phase; and modeling response/non-response in the second
phase is the final phase. More importantly, a design-unbiased variance
estimator for estimator
 (\ref{est}) is given in Theorem \ref{maintheorem}.  The estimator (\ref{variance2}) of  $Var(\hat{T_2})$ was developed by applying our three-phase variance estimator (\ref{varest}).

Although the focus here has been on estimating the number of US farms, the same ALUS follow-on and adjustment for non-response in the second phase allow unbiased estimates of other variables to also be obtained. The experience gained from the FNRP described in Section 2, the change in JAS protocols following the FNRP, and the fact that the FNRP included only 2009 segments could lead to the ALUS results being different from those anticipated here. ALUS has been proposed during a time of declining budgets, and its additional expense is the primary reason NASS has yet to implement ALUS.

Following the FNRP, additional training on JAS pre-screening was conducted,
and the time field enumerators were given to complete pre-screening was extended from two to four weeks. This resulted in an initial increase in the estimated number of farms, using equation (1), and then the estimates began to decrease. Some of the decrease may be due to a decline in the number of farms; however, misclassification may again be increasing. Currently, NASS is using modeling approaches to adjust for this misclassification in JAS. It is hoped that ALUS can be conducted at least once, allowing the estimates based on the methods presented here to be compared to the modeled results.

{\bf {\large Acknowledgements}}\\
The authors would like to thank the Editor, the Associate Editor and the referees for numerous suggestions that improved the presentation of this paper.\\
This work was supported by the National Agricultural Statistics Service, United States Department of Agriculture and the National Institute of Statistical Sciences. Hailin Sang was at the National Institute of Statistical Sciences, Pam Arroway was on faculty at North Carolina State University,  Kenneth K. Lopiano and Linda J. Young were at the University of Florida when most of this research was conducted.

  \vspace{0.4in}
  
{\bf \large Appendix}

{\bf Proof of Theorem \ref{maintheorem}}\\
The proof is an application of the variance formula $Var(X)=Var[E(X|Y)]+E[Var(X|Y)]$. We sketch the necessary steps for readers convenience. 

Recall $T=\sum_U y_k$ is the population total. From the design, it is easy to see that $\hat{t}_{\#}$ is unbiased for $T$. To provide the variance formula for this estimator, first decompose $\hat{t}_{\#}-T$  as
$$\hat{t}_{\#}-T=(\sum_S\breve {y}_k-\sum_Uy_k)+(\sum_R\breve {y}_k^* -\sum_S\breve{y}_k)+(\sum_F\breve {y}_k^{\#} -\sum_R\breve {y}_k^* )=A_S+B_R+C_F.$$
Now let $E_S(\cdot)=E(\cdot |S)$ and  $Var_S(\cdot)=Var(\cdot|S$) refer, respectively, to the conditional expectation and variance in phase two, given the outcome $S$ of phase one. We also define $E_R(\cdot)=E(\cdot|R)$ and $Var_R(\cdot)=Var(\cdot|R)$ similarly. 
Then, the variance of the three-phase estimator is
\begin{equation}\label{variance}
Var(\hat{t}_{\#})=Var(\hat{t}_{\#}-T)=Var[E(\hat{t}_{\#}-T|S)]+E[Var(\hat{t}_{\#}-T|S)].
\end{equation}
Given the first phase sample, $A_S$ is constant, and the second and third phase estimators are unbiased. Therefore,  
\begin{equation}\label{conExp1}
E(\hat{t}_{\#}-T|S)=E(A_S+B_R+C_F |S)=A_S+0+0=A_S.
\end{equation}
Since 
\begin{equation}\label{conVar1}
Var(\hat{t}_{\#}-T|S)=Var_S [E(\hat{t}_{\#}-T|R)]+E_S [Var(\hat{t}_{\#}-T|R)],
\end{equation}
by a similar argument as in (\ref{conExp1}), one can easily have 
\begin{equation}\label{conVar4}
Var(\hat{t}_{\#}-T|S)=Var_S (B_R )+E_S [Var(C_F|R)].
\end{equation}
From (\ref{variance}), (\ref{conExp1}) and (\ref{conVar4}),  
\begin{eqnarray}\label{conVar5}
Var(\hat{t}_{\#})&=&Var(A_S )+E\{Var_S (B_R )+E_S [Var(C_F|R)]\}\notag\\
&=&Var(A_S )+E[Var_S (B_R)]+E\{E_S[Var_R (C_F )]\}.
\end{eqnarray}
Here,
\begin{equation}\label{conVar6}
Var(A_S )=\sum\sum_U\Delta_{akp} \breve{y}_k \breve {y}_p,
\end{equation}
\begin{equation}\label{conVar7}
Var_S(B_R)=\sum\sum_S\Delta_{kp|S} \breve{y}_k^*\breve{y}_p^*,
\end{equation}
\begin{equation}\label{conVar8}
Var(C_F |R)=Var_R (C_F )=\sum\sum_R\Delta_{kp|R} \breve{y}_k^{\#} \breve{y}_p^{\#}.
\end{equation}
But, this variance formula (\ref{conVar5}) can not be applied directly. Therefore, a design-unbiased estimator of the variance is needed. For arbitrary constant $c_{kp}$,
\begin{eqnarray}\label{conExp2}
E\{E_S [E(\sum\sum_F c_{kp}/\pi_{kp|R} |R)]\}=E[E_S(\sum\sum_Rc_{kp})]\notag\\   
=E(\sum\sum_S\pi_{kp|S} c_{kp})=\sum\sum_U\pi_{akp}\pi_{kp|S} c_{kp}=\sum\sum_U\pi_{kp}^*  c_{kp}.
\end{eqnarray}                                                                    
Let $c_{kp}=\breve{\Delta}_{kp|S}^* \breve{y}_k \breve{y}_p$ in the above argument (\ref{conExp2}). A design-unbiased estimator of the first term of (\ref{conVar5}) is 
\begin{equation}\label{conVar5est}
\sum\sum_F\breve{\Delta}_{kp|S}^* \breve{y}_k \breve{y}_p/\pi_{kp|R}.                                                                                 \end{equation} 
Let $c_{kp}=\breve{\Delta}_{kp|S}\breve{y}_k^*\breve{y}_p^*$. By using the first two equalities of (\ref{conExp2}), a design-unbiased estimator of  $E[Var_S (B_R )]$ (the second term of (\ref{conVar5})) is 
\begin{equation}\label{expest1}
\sum\sum_F\breve{\Delta}_{kp|S} \breve{y}_k^* \breve{y}_p^*/\pi_{kp|R}.                                                                              \end{equation}
Let $c_{kp}=\Delta_{kp|R}\breve{y}_k^{\#}\breve{y}_p^{\#}$.  By using the first equality of (\ref{conExp2}), a design-unbiased estimator of the first term of $E\{E_S [Var_R(C_F)]\}$ (the third term of (\ref{conVar5})) is 
\begin{equation}\label{expest2}
\sum\sum_F\Delta_{kp|R}\breve{y}_k^{\#} \breve{y}_p^{\#}/\pi_{kp|R}. 
\end{equation}
Putting (\ref{conVar5est}), (\ref{expest1}) and (\ref{expest2}) together, we have (\ref{varest}), a design-unbiased estimator of (\ref{conVar5}).

\end{document}